\documentclass[aps,prb,twocolumn,superscriptaddress,showpacs,floatfix,amsmath,amssymb]{revtex4}
\usepackage{graphicx}
\usepackage{color}
\usepackage{xcolor,soul}
\usepackage{cancel}
\usepackage[normalem]{ulem}
\usepackage[scr=boondoxo]{mathalpha}

\DeclareMathAlphabet\mathbfcal{OMS}{cmsy}{b}{n}

\newcommand{\vm}[0]{ {\vec m} }
\newcommand{\bj}[0]{ {\bf j} }

\newcommand{\bA}[0]{ {\bf A} }

\newcommand{\vB}[0]{ {\vec B} }

\newcommand{\MGGA}[0]{ {\rm {\tiny \tiny MGGA} } }

\newcommand{\xc}[0]{ {\rm xc}  }

\begin{document}

\title{Physical spin torques  from exactly   constrained exchange-correlation torques}
\date{\today}

\author{Jacques K. Desmarais}
\email{jacqueskontak.desmarais@unito.it}
\affiliation{Dipartimento di Chimica, Universit\`{a} di Torino, via Giuria 5, 10125 Torino, Italy}

\author{Kamel Bencheikh}
\affiliation{Setif 1 University-Ferhat Abbas, Faculty of Sciences, Department of Physics and Laboratory of Quantum Physics and Dynamical Systems, Setif, Algeria}

\author{Giovanni Vignale}
\affiliation{Institute for Functional Intelligent Materials, National University of Singapore, 4 Science Drive 2, Singapore 117544}

\author{Stefano Pittalis}
\email{stefano.pittalis@cnr.it}
\affiliation{Istituto Nanoscienze, Consiglio Nazionale delle Ricerche, Via Campi 213A, I-41125 Modena, Italy}

\begin{abstract}
The problem of capturing  physical spin torques in non-collinear magnetic systems has dominated the scene of spin-density functional theory (SDFT) in
the last two decades. Progress has been hindered by the fact that the spin torque is directly connected to the divergence of the spin current, a quantity that is {\em extraneous} to SDFT --- thus leading to {\em spurious} exchange-correlation (xc) torques in the spin dynamics. Moreover, SDFT cannot rigorously include vector potentials and spin-orbit couplings. Here, we propose a solution that exploits the U(1)$\times$SU(2)-invariance of the xc energy of
SpinCurrent-DFT (SCDFT) --- an exact constraint that is not accessible to SDFT.
 Non-vanishing xc torques  obtained on non-collinear solutions are constrained by the  aforementioned exact internal symmetry and do not enter the propagation of the spin magnetization --- 
i.e., the spin dynamics involve {\em only} the physical currents and physical spin-torques.
\end{abstract}

\maketitle

{\em Introduction.}~ 
New phases of matter have emerged from the interplay between spin-orbit coupling (SOC) and low dimensionality, such as chiral spin textures and spin-polarized states, which are relevant to the generation and control of spin current and magnetization in spintronics\cite{Wolf2001,RevModPhys.76.323}. In spite of the rich physics and wide array of applications\cite{vasili2025molecular,picozzi2025spin,mellnik2014spin,fert2008nobel},
the spin magnetization $\vec{m}$ follows  a deceptively simple equation of motion\cite{Shi2006}
\begin{equation}
\frac{\partial \vec{{m}}}{\partial t}+\partial_\nu \vec{\mathcal{J}}_{\nu}  =  \vec{\mathcal{T}} \label{35}
\end{equation}
where the driving force  $\vec{\mathcal{T}}$ is the spin torque density, $\vec{\mathcal{J}}_{\nu}$ is the total spin current density and Einstein summation is implied over repeated indices\cite{noteGDFTnotation}.
Considering now a  vector potential ($A_\nu$), associated with the magnetic field ($\vec{B}$), and a spin-dependent vector potential ($\vec{A}_\nu$)
--- representing spin-orbit coupling (SOC) ---,  $\vec{\mathcal{T}}$  and $\vec{\mathcal{J}}_{\nu}$ have the expressions\cite{noteunits}
\begin{subequations}
\begin{equation}\label{eq:TandJ-1}
\vec{\mathcal{T}} = 2 (\vec{A}_\nu \times   \vec{\mathcal{J}}_{\nu} +    \vec{B} \times \vec{m}) 
\end{equation}
and
\begin{equation}\label{eq:TandJ-2}
\vec{\mathcal{J}}_{\nu}=\vec{J}_{\nu}+n \vec{A}_\nu + \vec{m} A_\nu
\end{equation}
\end{subequations}
with paramagnetic spin current density $\vec{J}_{\nu}$.\cite{noteCurrent} Eq. \eqref{eq:TandJ-1} and  Eq. \eqref{eq:TandJ-2} sharply point to the fact that  a calculation of $\vec{\mathcal{T}}$ requires a calculation of not only $n$ and $\vec{m}$, {\em but also} $\vec{J}_\nu$.

The  state-of-the-art approach, however, employs the single-particle  equations of Spin-DFT\cite{BarthHedin:72}
\begin{align}\label{KSSDFT}
\left[-\frac{1}{2}
\nabla^2+v_s + \vec B_s 
 \cdot \vec{\sigma}  \right]\Phi_k= \epsilon_k \Phi_k
\end{align}
where $\Phi_{k}$ are two-component Pauli spinors. 
These equations  {\em only} reproduce the interacting particle density, $n = \sum_{k} f_{k} \Phi_k^{\dagger} \Phi_k$,  and spin magnetization, $\vec{m} = \sum_{k} f_{k} \Phi_k^{\dagger} \vec{\sigma} \Phi_k$ (the ${f_{k}}$ are occupation numbers). 
The effective potentials are 
$v_s = v +v_{\rm H} + v_{\rm xc} = v + v_{\rm Hxc}$,
where  $v_{\rm H}$ is the usual Hartree potential, and $\vec B_s=\vec B+\vec B_{\rm xc}$.
The ${\rm xc}$-potentials are functional derivatives of the 
${\rm xc}$-energy functional, $E_{\rm xc}[n,\vm]$, with respect to the corresponding conjugate densities, namely,
$v_{\rm xc} = \frac{\delta E_{\rm xc}[n,\vm]}{\delta n}$
and
$\vec B_{\rm xc} =  \frac{\delta E_{\rm xc}[n,\vm]}{\delta \vec m}$.\cite{notedifferentiability}

Stretching the domain of applicability of Eq. \eqref{KSSDFT}, a scalar and a spin-dependent vector potential, representing, respectively,  a magnetic field coupling to the orbital motion of the electrons and a spin-orbit coupling, are often added to the Spin-DFT procedure --- a fact which can only be justified {\em a posteriori}.

The Kohn-Sham (KS) equation of (the unmodified) Spin-DFT implies\cite{Capelle2001} 
\begin{equation}\label{eq:td-m}
\frac{\partial \vec{{m}}}{\partial t}+\partial_\nu  \vec{{J}}_{s,\nu}  =
2 \vec{B}_s \times \vec{{m}}  
\end{equation}
which involves a {\em spurious}  xc-torque,
\begin{align}\label{eqn:wt}
         \vB_{\rm xc} \times \vec{m} = \frac{1}{2} \partial_\nu \left[  \vec{J}_{s,\nu} - \vec{J}_{\nu} \right ]\;.
\end{align}
 due to the
{\em mismatch} between the divergence of the true paramagnetic spin-current $\vec{J}_{\nu}$ and that of the KS system, $\vec{J}_{s,\nu}$. Eq. \eqref{eqn:wt}  shows that $\vec{B}_\xc $ is not collinear to $\vec{m}$. In and of itself this is not problematic, except that Eq. \eqref{eqn:wt} does not provide us with an exact constraint for devising approximations. 
In practice, the only useful information from Eq. \eqref{eqn:wt} is a global condition 
\begin{align}\label{eqn:zttsdft}
 \int d^3r \,   \vm  \times \vB_{\rm xc} = \vec{0}\;
\end{align}
which does not constrain $\vB_{\rm xc}$ at each point in space.

Correspondingly, the quest for finding properly constrained xc fields that generate an accurate xc torque has dominated the scene of non-collinear Spin-DFT in the last two decades\cite{Sharma2007a,ScalmaniFrisch:12,Bulik2013,eich2013transverse,eich2013trans2,Ullrich2018,Dewhurst2018,petrone2018efficient,goings2018current,komorovsky2019four,desmarais2021spin2,Hill2023,Pu2023,tancogne2023,Moore2025,huebsch2025}. We argue below that the resolution of the problem requires switching to Spin-Current-DFT (SCDFT). In this way, not only is the Spin-DFT methodology seamlessly extended to  account for vector potentials, SOC and spin currents {\em consistently,} but the task of dealing with xc-torques is resolved: namely spin dynamics {\em only} depends on the physical spin-torques and physical currents, while the ``spurious" xc torques only contribute  to the correct determination of the non-collinear ground-state solutions. 
 Crucially, the requirement of  local U(1)$\times$SU(2)-invariance of the xc-energy functional  is a powerful guide to the construction of an approximate xc functional.

{\em Spin-Current-DFT (SCDFT) -- the basics}.~
In SCDFT, the paramagnetic spin current (as well as the regular current) is a legitimate basic variable, also when electromagnetic couplings and SOC are considered. 
Thus, the
  KS equations
have the form~\cite{VignaleRasolt:87,VignaleRasolt:88,Bencheikh:03}
\begin{subequations}
\begin{equation}\label{KSeq}
\left\{ \frac{1}{2}\left( -i {\nabla} +   \mathbfcal{A}   \right)^2 + \mathcal{\widetilde{V}}  + \hat{V}_{\rm Hxc}\right\} \Phi_k = \epsilon_k \Phi_k,
\end{equation}
where $\mathbfcal{A}  = \bA +  \vec{\bA}  \cdot \vec{\sigma}$,
$\mathcal{\widetilde{V}}=\mathcal{V}-\frac{1}{2} \vec{A}_\nu \cdot \vec{A}_\nu$,\cite{noteGDFTsquare}
\begin{equation}
\hat{V}_{\rm Hxc} \equiv
 {\mathcal V}_{\rm Hxc} 
- \frac{i}{2} \left[ \partial_\nu \mathcal{A}_{{\rm xc},\nu}  + \mathcal{A}_{{\rm xc},\nu} \partial_\nu \right]\;,
\end{equation}
\end{subequations}
 $\mathcal{V}_{({\rm Hxc)}} = v_{({\rm Hxc})} + \vec{B}_{({\rm xc})} \cdot \vec{\sigma}$, $\mathcal{A}_{{\rm xc},\nu}=A_{{\rm xc},\nu} +  \vec{A}_{{\rm xc},\nu}  \cdot \vec{\sigma}$.
 The effective potentials are derived from the xc energy functional $E_{xc}[n,\vec m,j_\nu,\vec J_\nu]$ --- in particular,
$A_{{\rm xc},\nu}= \delta E_{\rm xc}/\delta j_\nu$
and
$
\vec{A}_{{\rm xc},\nu} = \delta E_{\rm xc}/\delta \vec{J}_\nu
$.
The current and the spin-current densities are given by
$
j_\nu = \frac{1}{2i} \sum_{k} {f_k} \left\{ \Phi^\dagger_k \left[ \partial_\nu \Phi_k\right] -  \left[ \partial_\nu \Phi^\dagger_k \right] \Phi_k \right\}$
and ${\vec{J}}_\nu = \frac{1}{2i} \sum_{k} {f_k} \left\{ \Phi^\dagger_k \vec{\sigma}  \left[ \partial_\nu \Phi_k \right] -  \left[ \partial_\nu \Phi^\dagger_k \right] \vec{\sigma} \Phi_k \right\}
$, respectively. 

$E_{\rm xc}[n,\vec m,j_\nu,\vec J_\nu]$ has a remarkable exact property --- local U(1)$\times$SU(2)  invariance ~\cite{VignaleRasolt:88,Bencheikh:03}. In detail, let us  consider  the following transformations.
A local U(1) transformation, is defined by
$  \Phi \rightarrow    \exp\left[ i  \chi \right] \Phi $
where $   \chi $ is a scalar function of the position.
A local SU(2) transformations is  defined by
$  \Phi \rightarrow \exp\left[ i {\vec{\lambda} \cdot \vec{\sigma} } \right]  \Phi $
where $\vec{\lambda}$ is a  vector function of the position. The overall effect of the above transformations is  to rotate the spinors and each of the phases of their components as well.  These position-dependent  transformations implement a continuum of {\em local} rotations.
The two ``rotations'' are performed conjointly in U(1)$\times$SU(2) transformations. 
With the exception of the particle density $n$, which is invariant under U(1)$\times$SU(2), all the other variables of SCDFT are transformed according to precise rules [See SM, Sec. I, Eqs. S3 and S6 for the expressions of the infinitesimal transformations].
The spin magnetization is rotated in spin space, the current density is ``translated" by a gauge-dependent term, and the spin current density is both rotated and translated. 

It is a fundamental result of SCDFT\cite{VignaleRasolt:88,Bencheikh:03} that the xc energy functional is invariant under these transformations, i.e.,
\begin{equation}\label{EXCInvariance}
E_{\rm xc}[n',\vec m',j_\nu',\vec J_\nu']=E_{\rm xc}[n,\vec m,j_\nu,\vec J_\nu]\,,
\end{equation} 
where the primed variables are connected to the unprimed variables by the U(1)$\times$SU(2) transformation (notice that $n$ is invariant, namely $n'=n$). 
This is a remarkable symmetry that holds for the xc functional, even as SOC breaks this symmetry for the full Hamiltonian.
While the symmetry of the exchange functional trivially reflects  the
U(1)$\times$SU(2) invariance of the Coulomb interaction (even in Spin-DFT),
 the U(1)$\times$SU(2) invariance of the correlation functional, instead, rests on the ability of the KS system to reproduce $n$, $\vec m$, $j_\nu$ and $\vec{J}_\nu$ (a condition which, in general, {\em cannot be enforced in Spin-DFT}).

While the importance of enforcing U(1) invariance in the xc-energy functional approximations has long been recognized,   delivering major improvements\cite{Dobson93,Becke-j02,TP05,Pittalis2006,Pittalis07,Rasanen09,bates2012harnessing,Tellgren2012:062506,furness2015current,Raimbault2015,Berger2015,Raimbault2016,tellgren2018uniform,Laestadius2014:782,laestadius2019kohn,pemberton2022revealing,richter2023meta},
the conjoint U(1)$\times$SU(2) invariance has been considered only occasionally. We argue below that progress in capturing proper xc torques and noncollinear solutions has correspondingly lagged behind.

From the equation of motion of the particle density and spin magnetization,
both for the interacting system and KS system, in view of the fact that all  the fundamental relevant densities are
reproduced, one readily derives~\cite{VignaleRasolt:88,Bencheikh:03}
\begin{subequations}
\begin{equation}\label{c1_2}
\partial_\nu   \left[ n  {A}_{{\rm xc},\nu}   + {\vm \cdot  \vec{A}_{{\rm xc},\nu}  } \right] = 0\;,
\end{equation}
and
\begin{align}\label{c2_2}
2 \left[\vec{B}_{\rm xc}\times \vec{m} +  \vec{A}_{{\rm xc},\nu} \times \vec{J}_\nu \right]=\partial_\nu\left[n\vec{A}_{{\rm xc},\nu}+ \vec{m} A_{{\rm xc},\nu} \right]\;.
\end{align}
\end{subequations}
In particular, Eq. \eqref{c2_2} 
shows that the torque of $\vec{A}_{{\rm xc},\nu}$ on $ \vec{J}_\nu$ and the torque of $\vec{B}_{\rm xc} $ on $\vec{m}$  are balanced by the divergence of an ``xc diamagnetic spin current", about which more will be said below.
Importantly, these equations can also be derived directly from the U(1) $\times$ SU(2) invariance of $E_{\rm xc}$.
In fact, focusing on ground states, it has been proved that  
Eq. \eqref{c1_2} and Eq. \eqref{c2_2} are satisfied by any xc-energy functional that is
U(1)$\times$SU(2)  invariant, i.e., satisfies Eq.~(\ref{EXCInvariance})~\cite{VignaleRasolt:88,Pittalis2017}. This invariance is a key exact condition to be satisfied for any xc functional approximation to be admissible.

For example, the  local spin density approximation (LSDA) can be applied straightforwardly to non-collinear states by aligning the $z$-axis with the local direction defined by $\vm$ \cite{KueblerWilliams:88,NordstroemSingh:96,OdaCar:98,Simoni2025:026108}. 
One can readily verify that this approximation  is U(1)$\times$SU(2) invariant and satisfies the xc-torque equations
of SpinCurrent-DFT, although with vanishing  xc-torque. 
Therefore, the LSDA is an
admissible starting point for SpinCurrent-DFT. It is however a  very crude approximation, as
it does {\em not} account for the  longitudinal and transverse spin gradients and does not include the functional dependence of $E_{xc}$ on spin currents.

Going beyond the LSDA while preserving U(1) $\times$ SU(2) invariance is quite a challenging proposition.  
Several attempts have been made
to include \emph{transverse} gradients of the  magnetization\cite{landau2013electrodynamics}
at the level of the generalized-gradient approximations (GGAs)\cite{ScalmaniFrisch:12,Bulik2013,sharma2018source,Pu2023,Moore2025}. 
Unfortunately, these attempts break U(1)$\times$SU(2) invariance and, thus, are not admissible in SpinCurrent-DFT\cite{noteGB}.

The situation has changed in recent years with the advent of U(1)$\times$SU(2)-invariant xc functionals of the Meta-Generalized Gradient Approximation (MGGA) class, which use the kinetic energy density and its spin-dependent generalization as building blocks~\cite{Pittalis2017,boccuni2024unveiling,DesmaraisSCMGGA,DesmaraisNCmSCAN}. Because these building blocks are naturally expressed in terms of single-particle spinors, they are most efficiently handled in the framework of the generalized Kohn-Sham (GKS) theory --- a broadening of the traditional KS theory in which the minimization of the energy with respect to densities is replaced by a minimization with respect to single-particle states, in terms of which the densities are expressed.~\cite{Seidl1996,DesmaraisGKS,Nrep}

{\em Spin-Current-DFT --- unleashed}.~ 
MGGAs are yielding major improvements in modern DFT~\cite{SCAN,CHEMSCAN,LAK,SKALA}.
They represent the xc energy functional as the spatial integral of an xc energy density of the form ${\epsilon}_{\rm xc}(n,\nabla n, \tau$) 
where $\epsilon_{\rm xc}$ is a regular function of three  arguments 
where 
$n =  \sum_i f_i |\varphi_i|^2$ and  $\tau = \frac{1}{2}\sum_i f_i|\nabla \varphi_i|^2$ are the usual (DFT) particle and kinetic energy densities. MGGAs can also depend on $\nabla^2 n$, yielding to no substantial difference in the arguments that follow.

An analysis based on the short-range behavior of the exchange hole showed that  MGGAs can be turned into U(1)$\times$SU(2)-invariant approximations satisfying Eq.~(\ref{EXCInvariance}) 
via the minimal substitution~\cite{Pittalis2017}
 $\tau \rightarrow   \widetilde{\tau}  = \left(  \tau - \frac{j_\nu j_\nu }{ 2n }  \right)  + \left( \frac{\vec{m}  \cdot \vec{\tau}}{n}   - \frac{ \vec{J}_\nu \cdot \vec{J}_\nu  }{ 2n  }  \right)  +   \frac{ \partial_\nu \vec{m}  \cdot  \partial_\nu \vec{m} }{8 n }$ [See Sec. II in SM].\cite{notetau}
We stress that the particle-kinetic energy density
$\tau = \frac{1}{2} \sum_k f_k \Big(  \partial_\nu \Phi^{\dagger}_k \Big) \Big( \partial_\nu \Phi_k \Big)$ and  the spin-kinetic energy density
$
{\vec{\tau}} =  \frac{1}{2} \sum_k f_k \Big(  \partial_\nu \Phi^{\dagger}_k \Big) \vec{\sigma} \Big( \partial_\nu \Phi_k \Big)$
are   built from two-component (2c) spinors, and transverse  gradients of the spin magnetization are included in $( {\partial_\nu} \vec{m}  )  \cdot ( {\partial_\nu} \vec{m})$. 
The quantity $\widetilde{\tau}$ has recently emerged as the cornerstone of  the  U(1)$\times$SU(2)-invariant extension of the electron localization function to non-collinear spin states \cite{zwoelf}.  It also enters the U(1)$\times$SU(2)-invariant extension of the SCAN functional (the non-collinear-mSCAN, or briefly mSCAN)\cite{DesmaraisNCmSCAN}, which fixes serious shortcomings  of the original SCAN functional
for magnetic materials.
\cite{noteGDFTmin} For benchmarks of the non-collinear mSCAN functional on spin splitting in molecules and materials, and for spin-moments, energies and energy gradients, as well as dissociation of diatomic molecules we refer the reader to Refs. \onlinecite{DesmaraisSCMGGA,boccuni2024unveiling,DesmaraisNCmSCAN}.

MGGAs find their most comfortable home in DFT via the GKS formalism\cite{Seidl1996,perdew2017understanding}. The minimization of the energy functional is carried out with respect to the spin-orbitals, with the  densities regarded as functions of the spin-orbitals. This is at least as general as the standard KS procedure within the space of densities that are non-interacting $v$-representable\cite{Seidl1996,DesmaraisGKS}.

For U(1)$\times$SU(2) invariant MGGAs, in the presence of SOC and electromagnetic couplings the variational principle leads to the GKS equations in the form
\begin{subequations}
\begin{align}\label{GKS-SCDFT}
\left\{ \frac{1}{2}\left( -i {\nabla} +   \mathbfcal{A}  \right)^2 + \mathcal{\widetilde{V}}  + \hat{V}^\MGGA_{\rm Hxc}\right\} \Phi_k = \epsilon_k \Phi_k,
\end{align}
where
\begin{equation}
\hat{V}^\MGGA_{\rm Hxc} \equiv
 \mathcal{V}_{\rm Hxc}  - \frac{1}{2}\partial_\nu  {\cal M }_{{\rm xc}}  \partial_\nu 
- \frac{i}{2} \left[ \partial_\nu \mathcal{A}_{{\rm xc},\nu}  + \mathcal{A}_{{\rm xc},\nu} \partial_\nu \right]\;
\end{equation}
\end{subequations}
with
 ${\cal M }_{\rm xc}= M_{\rm xc} + \vec{\sigma} \cdot \vec{M}_{\rm xc}$. Notice that
${\cal M }_{\rm xc}$ can be understood as an effective (inverse) mass accounting for both an Abelian (i.e. particle), $M_{\rm xc} =\partial \epsilon_{\rm xc} / \partial \tau$, and  a non-Abelian  (i.e., spin), $\vec{M}_{\rm xc} =\partial \epsilon_{\rm xc} / \partial \vec{\tau}$, component.

From Eq. \eqref{GKS-SCDFT}, it is apparent that the GKS equation replaces the several {\em local} potentials of  the regular KS-SCDFT [see Eq. \eqref{KSeq}]  by
a {\em single} semi-local operator $\hat{V}^\MGGA_{\rm xc}$. Yet, $\hat{V}^\MGGA_{\rm xc}$ is obtained in terms 
of several {\em local} fields: $v_{\rm xc}$,  $\vec{B}_{\rm xc}$, $M_{\rm xc}$, $\vec{M}_{\rm xc}$, $A_{{\rm xc},\nu}$, $\vec{A}_{{\rm xc},\nu}$. In detail, for a MGGA as the nc-mSCAN --- with energy density per unit volume $\epsilon_{\rm xc} = {\epsilon}_{\rm xc}(n,\partial_\nu n, \vert \vec{m} \vert, \widetilde{\tau}(n, \vec{m},\partial_\nu \vec{m},j_\nu,\vec{J}_\nu,\tau,\vec{\tau}))$ --- one gets 
the explicit expressions reported in Sec. III of the SM, Eqs. (S24)-(S28).

Next, {\em Eqs.~\eqref{c1_2} and \eqref{c2_2} must also be generalized}. Crucially, from the U(1) invariance [see Sec. IV in SM] 
and the SU(2) invariance of the functional [see Sec. V in SM] plus the remarkable identities in Sec. VI of the SM,
we obtain
\begin{subequations}\label{eqn:newteqs}
\begin{align}\label{eqn:newteqs-1}
     \vm \times \vec{B}_{\rm xc} +\vec{\tau} \times   \vec{M}_{\rm xc}  =   \frac{1}{4} \partial_\nu \left[ \left( \partial_\nu \vec{m} \right) \times  \vec{M}_{\rm xc}  \right]\;. 
\end{align}
Integration, readily, implies
\begin{align}\label{eqn:newteqs-2}
\int d^3r ~ 
 \left[ \vm \times \vec{B}_{\rm xc} +\vec{\tau} \times   \vec{M}_{\rm xc} \right]
= \vec{0}\;.
\end{align}
\end{subequations}
The latter  relation can also be derived by  generalization of the zero-torque theorem [see
Sec. VII in SM].

Eq.s \eqref{eqn:newteqs-1} and \eqref{eqn:newteqs-2} upgrade  Eq. \eqref{eqn:wt} and Eq. \eqref{eqn:zttsdft} to a larger domain of applicability. 
In particular, Eq. \eqref{eqn:newteqs-1} and Eq. \eqref{eqn:newteqs-2} include an additional xc-torque (i.e., $\vec{\tau} \times   \vec{M}_{\rm xc} $) and  the right hand side of Eq. \eqref{eqn:newteqs-1}, that is, $1/4 \partial_\nu \left[ \left( \partial_\nu \vec{m} \right) \times  \vec{M}_{\rm xc}  \right]$,
suggestively resembles the divergence of the  {\em phenomenological} ``spin currents'' of the Landau-Lifshitz equation \cite{maekawa2017spin}. 
To the best of our knowledge, these terms have been missed or gone unnoticed in  previous MGGA-related works on non-collinear magnetism.

The remarkable identities in Sec. VI of the SM,  imply that the total (i.e. paramagnetic plus diamagnetic) GKS  particle- and spin-currents are --- in form --- identical to their exact interacting counterparts. Next, we invoke the adiabatic {\em approximation} --- i.e.,  all the densities are derived by solving the GKS equations in time and the  instantaneous 
spinors are used as an input to a ground state functional [see Sec. VIII in SM]. As a result,  the {\em adiabatic} application of the proposed MGGAs yields an equation of motion for the
 particle density 
 \begin{subequations}
 \begin{equation}\label{22}
\frac{\partial n}{\partial t} +\partial_\nu \mathscr{j}_{\nu}=0 
\end{equation}
where the total particle current is
\begin{equation}
\mathscr{j}_\nu=j_\nu+nA_\nu + \vec{m} \cdot \vec{A}_\nu
\end{equation}
\end{subequations}
and yields Eq. \eqref{35} for the propagation of the spin magnetization. These
have the same form as in the (exact) interacting case. 
In other words, the time-evolution of  both $n$ and $\vec{m}$ are free from the action of spurious  xc-torques (even though non-zero xc-torques {\em do} implicitly occur in the solutions). This provides a substantial extension of previous works on the generalized continuity equation for the particle density\cite{bates2012harnessing,Baer18,richter2023meta} --- here extended to the spin magnetization.

\begin{figure}[h!]
\label{fig:Crmono}
\centering
\includegraphics[width=8.6cm]{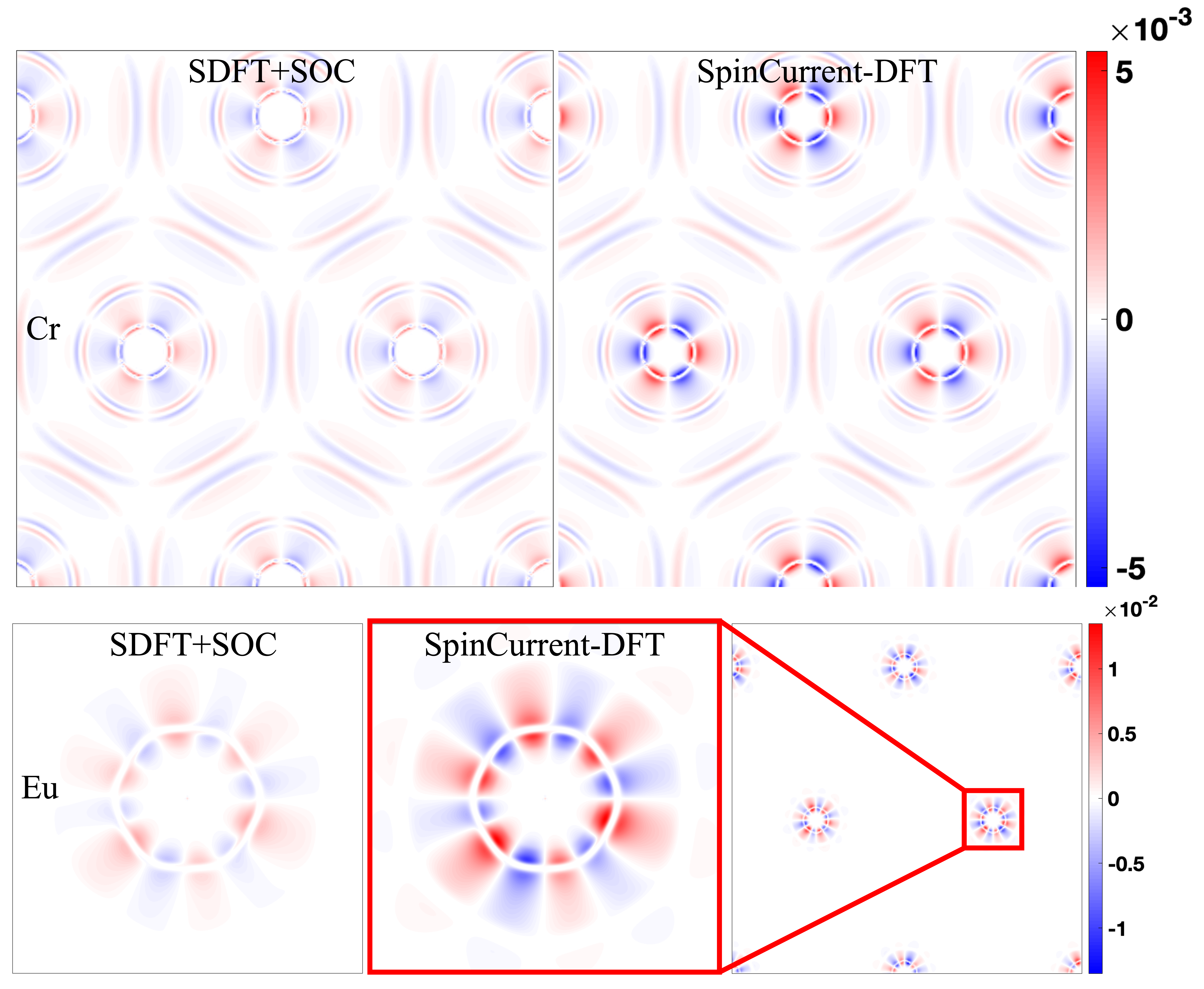}
\caption{Plots of the xc torques for (top) the Cr and (bottom) Eu monolayers due to the noncollinear mSCAN functional in a (left) Spin-DFT+SOC $\vec{m}\times \vec{B}_{\rm xc}$ and (right) SpinCurrent-DFT $\vec{m}\times \vec{B}_{\rm xc}+\vec{\tau}\times\vec{M}_{\rm xc}$ formulation. Colors denote magnitude in a.u.} 
\end{figure}

{\em Further analyses and applications}:~ 
To make contact with previous work on the xc torque in Spin-DFT,\cite{Sharma2007a,ScalmaniFrisch:12,Bulik2013,eich2013transverse,eich2013trans2,Ullrich2018,Dewhurst2018,goings2018current,komorovsky2019four,desmarais2021spin2,Hill2023,Pu2023,tancogne2023,Moore2025,huebsch2025}  we plot the key quantities appearing in Eqs. \eqref{eqn:newteqs-1} and  \eqref{eqn:newteqs-2}. A standard system on which such plots are usually reported is the unsupported Cr (001) monolayer in the N{\'e}el state.\cite{SharmaGross:07,bulik2013noncollinear} To also provide an example in a stronger SOC regime, we additionally consider the analogous monolayer with Cr atoms replaced by Eu.

We performed self-consistent SpinCurrent-DFT calculations with the MGGA (the non-collinear mSCAN functional),\cite{DesmaraisNCmSCAN} including SOC [details are provided in Sec. IX in SM]. As in previous works,\cite{SharmaGross:07,bulik2013noncollinear} we employed a lattice parameter of 5 \AA, which leads to geometrical frustration of the monolayers and, correspondingly, a two-dimensional non-collinear spin wave. 
We compare an empirical Spin-DFT+SOC against SpinCurrent-DFT implemented with the same MGGA. Within the present approach, the empirical  Spin-DFT+SOC scheme is obtained by neglecting the dependence of the xc functional on currents (thus breaking the gauge invariance of the functional)--- as done, for instance, in Ref. \onlinecite{tancogne2023}. Results for the xc-torque component normal to the (001) planes are presented in the contour plots of Fig. 1 (top) for Cr and (bottom) for Eu. The xc-torque from Spin-DFT+SOC is visibly different to the xc-torque in SpinCurrent-DFT. In Eu, the Spin-DFT+SOC result has around half the SpinCurrent-DFT value. Differences between Spin-DFT+SOC and SpinCurrent-DFT on the total xc-torque are also visible for Cr, though the effect is smaller than in Eu.

First of all, for the exact condition provided in Eq. \eqref{eqn:newteqs-2}, we obtained $\int d^3r \  \vec{m} \times   \vec{B}_{\rm xc}= \vec{0}$, as well as $\int d^3r \  \vec{\tau} \times   \vec{M}_{\rm xc}= \vec{0}$ for the Cr (001) and Eu (001) monolayers. In general, however, from Eq. \eqref{eqn:newteqs-2}, we expect that only the total xc torque should integrate to zero, even if the integrals of the individual torques are not zero.

To quickly discuss an example with non-zero integrated quantities $\int d^3r \  \vec{\tau} \times   \vec{M}_{\rm xc} \ne 0$ and $\int d^3r \  \vec{m} \times   \vec{B}_{\rm xc} \ne \vec{0}$, we also performed analogous calculations on a noncoplanar magnet which was built from the Cr$_4$ cluster in a tetrahedral arrangement (geometry in Sec. IX in SM). This provided an integrated $\Vert \int d^3r \  \vec{\tau} \times   \vec{M}_{\rm xc} \Vert$ (and $\Vert \int d^3r \  \vec{m} \times   \vec{B}_{\rm xc} \Vert$) of +(-) 6.82 $\times$10$^{-4}$ a.u.  A discussion on the contribution of individual terms $\vec{m} \times   \vec{B}_{\rm xc}$ and $\vec{\tau} \times   \vec{M}_{\rm xc}$ also to the plot of the xc-torque is provided in Sec. X of the SM.

Finally, it is particularly interesting to discuss the effect of SOC on the xc-torques more specifically. Fig. 2 for (top) Cr and (bottom) Eu provide the differences of xc-torque due to SOC, both in Spin-DFT+SOC and SpinCurrent-DFT. For the considered systems, we find that the effect of SOC on the xc-torques in Spin-DFT+SOC are grossly underestimated compared to the effect observed in SpinCurrent-DFT, irregardless of the SOC intensity. {For the case of Eu, the difference due to SOC is on the same order of magnitude as the total in SpinCurrent-DFT, but almost negligible in Spin-DFT+SOC} (compare bottom panels of Figs. 1 and 2).

\begin{center}
\begin{figure}[h!]
\label{fig:delta}
\centering
\includegraphics[width=8.6cm]{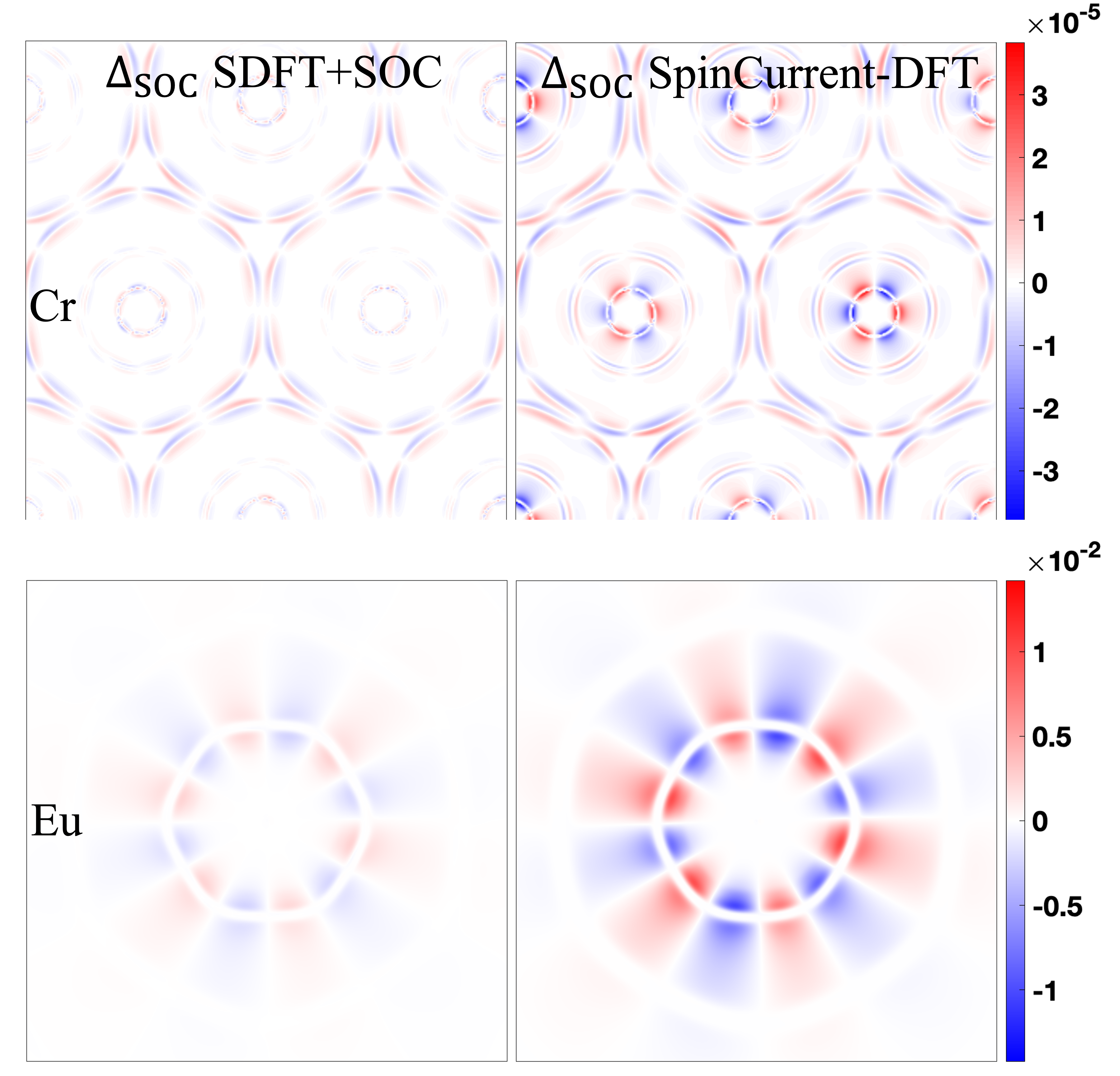}
\caption{Plots of the {xc-torque differences} for (top) the Cr and (bottom) Eu monolayers due to SOC for Spin-DFT+SOC and for SpinCurrent-DFT.} 
\end{figure}
\end{center}

{\em Conclusions}.~
Difficulties in capturing exchange-correlation (xc)-torques non empirically, have  persisted in Spin-DFT (SDFT). 
Strictly speaking, spin-torques involving spin-currents, vector potentials, and spin-orbit couplings are beyond the reach of SDFT.
Here, we have presented a solution based  on first principles, which extends both Spin-DFT and the early formulation of SpinCurrent-DFT.

The newly derived Eq. \eqref{eqn:newteqs-1} and Eq. \eqref{eqn:newteqs-2}  involve  xc-torques
due to the non-collinearity of the targeted states. These torques  follow from a U(1)$\times$SU(2) invariant extension of meta-generalized-gradient approximations of SpinCurrent-DFT. 
Eq. \eqref{eqn:newteqs-1} and Eq. \eqref{eqn:newteqs-2} upgrade Eq. \eqref{eqn:wt} and Eq. \eqref{eqn:zttsdft} to a larger domain of applicability than allowed by present-day Spin-DFT.  In fact, the U(1)$\times$SU(2) invariance of the xc-energy functional is an exact {\em local} constraint that, together with a consistent account of spin-orbit coupling, is {\em not} available in SDFT-based approaches.

To make contact with previous results, we have analyzed  the novel xc-torques in applications, finding (as expected)  significant differences w.r.t. a Spin-DFT based approach when  spin-orbit coupling is important.

Crucially, we have shown that the derived xc-torques are such that the {\em adiabatic} time propagation reproduces --- in form --- the   dynamics of the interacting particle density and  spin magnetization: the xc torques  only contribute to the determination of the non-collinear ground state solutions and not explicitly to the  physical spin torque.

The  results presented in this work point to SpinCurrent-DFT as an ideal framework for  tackling systems at equilibrium and out-of-equilibrium including    couplings that are beyond Spin-DFT.

\begin{acknowledgments}
JKD was supported by the Project CH4.0 under the ``Ministero dell’Universit\`a e della Ricerca'' (MUR) program ``Dipartimenti di Eccellenza 2023-2027'' (CUP: D13C22003520001). GV was supported by the Ministry of Education, Singapore, under its Research Centre of Excellence award to the Institute for Functional Intelligent Materials (I-FIM, project No. EDUNC-33-18-279-V12). SP was partially supported by  MUR under the Project PRIN 2022 number 2022W9W423.
\end{acknowledgments}


\begin{thebibliography}{87}
\bibitem[{not({\natexlab{a}})}]{noteGDFTnotation}
\bibinfo{note}{Vectors with a spin index (e.g., ``$a$'') may be denoted with an
  arrow (e.g., $\vec{m}$). Vector carrying a spatial index (e.g., ``$\nu$'')
  may be denoted with a bold character (e.g., $\bj$). A tensor carrying both
  spin and spatial index can be denoted as a spatial vector with a spin index
  (e.g., $\bA^a$) or a spin vector with a spatial index (e.g., $\vec{A}_\nu$),
  as it may be more convenient. {In particular, it should be borne in mind that
  $A_{\nu}$ repesents the $\nu$ component of a spatial vector --- not the
  moduli of $\vec{A}_{\nu}$. }}.
\bibitem[{not({\natexlab{c}})}]{noteCurrent}
\bibinfo{note}{The paramagnetic current differs from the full current in that
  the latter is constructed from the velocity operator (the time derivative of
  the position operator) while the former employs the ``canonical velocity"
  ${\bf p}/m$ where ${\bf p}$ is the momentum operator.}
\bibitem[{not({\natexlab{e}})}]{noteGDFTsquare}
\bibinfo{note}{In writing the minimal substitution for the non-Abelian vector
  potential representing SOC, we have taken into account that the physical case
  of interest in this work includes only linear terms. Therefore, as usual,
  $1/2 \vec{A}_\nu \cdot \vec{A}_\nu$ is substracted out and grouped with
  $\mathcal{V}$. Importantly, the gauge invariance of the xc-energy functional
  is preserved by the fact that the external potentials do not enter the
  definition of the xc functional explicitly.}
\bibitem[{not({\natexlab{g}})}]{notetau}
\bibinfo{note}{Please note that $\tau$, $\vec{\tau}$ and $\tilde{\tau}$ are
  three distinct quantities.}
\end{thebibliography}
\end{document}